\documentclass[12pt]{article}
\begin{document}
\def\thebibliography#1{\section*{REFERENCES\markboth
 {REFERENCES}{REFERENCES}}\list
 {[\arabic{enumi}]}{\settowidth\labelwidth{[#1]}\leftmargin\labelwidth
 \advance\leftmargin\labelsep
 \usecounter{enumi}}
 \def\newblock{\hskip .11em plus .33em minus -.07em}
 \sloppy
 \sfcode`\.=1000\relax}
\let\endthebibliography=\endlist

\hoffset = -1truecm
\voffset = -2truecm

\title{\large\bf
Can Environmental Decoherence be Reversed for an Open Quantum System
in a Magnetic field ?}
\author{
{\normalsize \bf A.N. Mitra$^\star$ \thanks{e.mail:
ganmitra@nde.vsnl.net.in} }
\\
\normalsize  Department of Physics, University of Delhi,\\
Delhi-110007, India  \\
}
\date{}
\maketitle


\begin{abstract}A simple model is considered  for an open system consisting of an
aggregation of magnetic particles (like greigite) in the presence of a
magnetic field (H), and interacting linearly with a bath of 3D
harmonic oscillators. Using the Feynman-Vernon
formalism, as given in Weiss (termed FVW),  the time-evolved
reduced density matrix ( after eliminating the bath d.o.f.'s), is
examined for environmental decoherence as defined in the FVW
formalism. While decoherence is usually positive for most two-way
couplings with the enviroment, it is found  that a $three-way$
interaction involving the system plus bath plus H-field all together, can
facilitate a $reversal$ of sign of this quantity ! This may have implications
for quantum coherence based phenomena on the origins of life.
\\
Key words :  Decoherence; field-induced aggregate; H-field; 3-way coupling; sign reversal \\
PACS : 03.65.Yz; 05.40.-a; 03.67.-a; 75.75.Jn \\
\end{abstract}

\section{Introduction : Preliminaries }

Decoherence is a ubiquitous phenomenon [1] which arises when one attempts to extend  quantum theory
  to macro systems (and even to meso systems ).  Since environment has
 a crucial role for a quantum  system, the relevant theoretical framework
 for the study of decoherence is a theory of open quantum systems which treats
 the effect of an uncontrollable environment on the quantum evolution.
 The concept was originally developed to incorporate the effect of friction
 and thermalization in a quantum formalism, but a correct perspective of
 decoherence effects requires its time scale to be much shorter than typical
 `relaxation'  phenomena. Rather ironically, decoherence is not so much about
 $how$ a bath d.o.f. affects the system, as $vice versa$, the latter revealing information
 on the state of the system.  The relevance of  decoherence is especially acute for quantum
 information  processing tasks where the
 coherence  of many quantum systems must be maintained for a long time.
To explain decoherence in a nut-shell, one may crudely assume that the
interaction between the quantum system and environment is short-ranged enough
to be described by scattering theory for asymptotically `free' states before
and after the interaction, so that the detailed anatomy of the collision dynamics
itself does not come into the picture. Let $\rho$ be density operator for the
system which, for simplicity,  interacts with a single environment d.o.f.
represented by $|\psi>$ . Then it can be shown that   the $diagonal$ elements
of $\rho$ remain unchange by interaction, i.e.,  $\rho'_{mm}$ = $\rho_{mm}$ . But
the off-diagonal elements $\rho'_{mn}$ after interaction are no longer equal
to $\rho_{mn}$ before interaction. Rather, they get suppressed by a factor
$<\psi^n |\psi^m>$ whose magnitude is always less than unity. And since the
off-diagonal elements are the repository of quantum effects, this suppression
is tantamount to loss of $ability$ of the system to show quantum behaviour due
to the interaction with the environment's  quantum d.o.f.  [It arises from the correlation
between the system plus environment particle]. So, if the environmental d.o.f. is
integrated out, this $ability$ is lost. Alternatively, the interaction is an information
transfer from the system to the environment. The more the overlap  $<\psi^n |\psi^m>$
deviates from unity,  the more an observer could in principle learn about the
system by measuring the environmental particle ! But this measurement is usually never
made, so this alternative principle explains that the wave-like interference phenomenon
characterized by the coherence $vanishes$  as more information discriminating the distinct
" particle-like"  system  is revealed.
\par
Against this background, we may regard that decoherence that is inherent in the decay of an
open quantum system into the surroundings, leads to a destruction
of phase correlations within the system. The issue of decoherence of
such a system due to its interaction with a thermal bath of harmonic
oscillators  was first formulated by Feynman and Vernon [2] and
addressed subsequently by many workers, notably
Grabert-Schramm-Ingold [3] as well as Caldiera-Leggett [4]. The
result of many of these studies has been incorporated in a
comprehensive book by Weiss [5] which will be taken as a standard
reference for the present investigation. We shall freely use the
results of ref [5],often under the name FVW, and in the same
notation where possible without explanation. Specifically we
consider an `open' system consisting of an aggregation of magnetic
particles (such as greigite) in the presence of a magnetic field (H),
and interacting linearly with a bath of 3D harmonic oscillators.
While several types of two-way interaction of the system with the
bath has been adequately covered in FVW, the possibility of new
forms of interactions when an external magnetic field is present,
and their effect on the resulting decoherence, is the subject of
this paper.
\par
In Section 2 we collect the relevant formulae from ref.[5] under the
further assumption of slow time variation, so as to simplify the
essential formulae, as well as extend them from 1 to 3 dimensions
before identifying the decoherence rate. In Section 3 we introduce
the H-field and examine the effect of a new form of interaction--a
$three-way$ coupling involving the system, bath and field. To the
best of our knowledge such a $direct$ 3-way coupling has not been considered in
the literature [5], a novel feature being  that the calculated
decoherence has a $negative$ sign; its origin is of course accounted
for, and its order of magnitude estimated. Section 4 discusses the
significance of this result and its possible interpretation on the
lines of a Froehlich-like  mechanism [6].(In Appendix A we re-derive
a formula for the problem 3-11 of Feynman-Hibbs [7], since a term
appears to be missing in it).

\section{Essential Results for System plus Bath Coupling}

\setcounter{equation}{0}
\renewcommand{\theequation}{2.\arabic{equation}}

From Eq (3.11) of Weiss[5], the System-Bath Hamiltonian has the form
\begin{equation}\label{2.1}
H = \frac{p^2}{2M}+ \Sigma_1^N \frac{p_\alpha^2}{2m_\alpha} + V(q,x)
\end{equation}
\begin{equation}\label{2.2}
V(q,x) = V(q) + \frac{1}{2}\Sigma_1^N m_\alpha
\omega_\alpha^2(x_\alpha - \frac{c_\alpha q}{m_\alpha
\omega_\alpha^2})^2
\end{equation}
This Hamiltonian serves two purposes : A) As the Euclidean
Lagrangian (with imaginary Matsubara time [5])in quantum statistical
mechanics for the calculation of the density matrix under thermal
equilibrium conditions a la Feynman-Hibbs [7] and Weiss [5]; B) as
the Hamiltonian corresponding to a real-time Lagrangian (obtained by changing
the  sign for the potential term V) which participates in the
time-evolution of the density matrix for the joint system-bath
complex under non-equilibrium conditions, using real time path
integral techniques [4, 5, 7]. The Matsubara time $\tau$ is related
to the real time t by $t = -i\tau$, and varies in the range
\begin{equation}\label{2.3}
 0 \leq \tau \leq \hbar \beta = \hbar/ kT
\end{equation}
The global density matrix $W(0)$  for the initial state t=0 is assumed for
simplicity to be in a factorized form in which the system and bath
are decoupled [4,5],and the unperturbed bath is in thermal equilibrium,
given by (see eq (5.10) of [5]):
\begin{equation}\label{2.4}
W(0) = \rho(0) \bigotimes W_R(0); \quad W_R(0) = Z_R^{-1}\exp\{-\beta H_R\}
\end{equation}
Here $\rho$ is the reduced density matrix for the system whose time
evolution $\rho(t)$ is the primary object of interest. To that end
note that the global density matrix $W(t)$ evolves from a state $W(0)$
at t=0 as (see eq.(5.1) of [5])
\begin{equation}\label{2.5}
W(t)= \exp\{-iHt/\hbar\} W(0) \exp\{+iHt/\hbar\}
\end{equation}
We now skip most of the formalism which is outlined in sections (5.1-2) of
Weiss [5], and focus on the time evolution $\rho(t)$ of the reduced
density matrix after eliminating the bath d.o.f.'s. This quantity can be
expressed in terms of the Feynman-Vernon influence functional $F_{FV}$
as follows [5]
\begin{equation}\label{2.6}
\rho(q_f q_f';t) = \int dq_i dq_i' \int Dq Dq' \exp[ i(S_S[q]-S_S[q'])/\hbar]F_{FV}[q,q']
\end{equation}
$F_{FV}$ in turn is made up as a product of two functionals $F, F^*$ :
\begin{equation}\label{2.7}
F_{FV}[q,q']= \int dx_f dx_i dx_i' W_R[x_i,x_i']F[q;x_f x_i]F^*[q';x_f x_i']
\end{equation}
where
\begin{equation}\label{2.8}
F[q;x_f x_i] = \int Dx(.)\exp[ i(S_R[x])+S_I[x,q])/\hbar]
\end{equation}
and a similar expression for the (complex conjugated) quantity $F^*[q';x_f x_i']$.
The quantity $W_R[x_i,x_i']$ in eq(2.7) which also appears in eq (2.4), is the
canonocal density matrix for the bath in thermal equilibrium, and its
path integral over the (imaginary) Matsubara time $\tau$ is (see eq (5.17) of [5])
a product of modes $W_{R\alpha}$  where (see eq (5.17) of [5])
\begin{equation}\label{2.9}
W_{R\alpha}[x_i,x_i'] = C_\alpha \exp\{ \frac{-m_\alpha \omega_\alpha}{2\hbar\sinh\beta\hbar\omega_\alpha}
[(x_{i\alpha}^2 + x_{i\alpha}'^2)\cosh\beta\hbar\omega_\alpha - 2x_{i\alpha}x_{i\alpha}']\}
\end{equation}
Similarly the quantity $F[q;x_f x_i]$ is expressible as a product of its modes $F_\alpha$
whose path integral over real time [7,5] works out as [5]
\begin{equation}\label{2.10}
F_\alpha [x_{f\alpha}, x_{i\alpha}] = C_\alpha(t)\exp\{\frac{i}{\hbar}\phi_\alpha[q;x_{f\alpha}, x_{i\alpha}]\}
\end{equation}
where $C_\alpha(t)$ is a constant, and  $\phi_\alpha$ is given by eq (5.19) of ref [5]. However we prefer to
give a simplified expression for $\phi_\alpha$ and a corresponding quantity $\phi_\alpha'$ arising from
$F^*[q';x_f x_i']$ of eq (2.7), based on an (approximate) time independence of $q$ and $q'$, which may be
adapted from eq.(A.7) of Appendix A. The result for $\phi_\alpha$, mostly in the notation of ref [5], is
\begin{eqnarray}\label{2.11}
\phi_\alpha&=& \frac{m_\alpha\omega_\alpha}{2\sin\omega_\alpha t}[(x_{i\alpha}^2 + x_{f\alpha}^2)\cos\omega_\alpha t
-2 x_{i\alpha}x_{f\alpha}] \nonumber  \\
           &+& c_\alpha (x_{i\alpha}+ x_{f\alpha})q \frac{1-\cos\omega_\alpha t}{\omega_\alpha \sin\omega_\alpha t}
- \frac{c_\alpha^2 q^2(1-\cos\omega_\alpha t)^2}{2 m_\alpha\omega_\alpha^3\sin\omega_\alpha t} \nonumber  \\
           &+&  \frac{c_\alpha^2 q^2}{2 m_\alpha\omega_\alpha^2}[t - \frac{\sin\omega_\alpha t}{\omega_\alpha}]
\end{eqnarray}
The corresponding quantity $\phi_\alpha'$ which comes with a negative sign, arising as it does from $F^*[q';x_f x_i']$,
is given by a similar expression to (2.11), except for
$$ x_{i\alpha} \Rightarrow x_{i\alpha}'; \quad q \Rightarrow q'$$

The rest of the calculation which involves the integration over the bath d.o.f.'s,  is straightforward though
lengthy. Following Weiss [5], introduce the variables
\begin{equation}\label{2.12}
\sqrt{2} [z_{i\alpha}; y_{i\alpha}]= x_{i\alpha} \pm  x_{i\alpha}'
\end{equation}
The integration in eq(2.6) for the reduced density matrix $\rho$, over the variables $z,y, x_f$, all standard
gaussian variables, is best carried out successively in this very order. The result for the related Feynman-Vernon
influence functional $F_{FV}[q,q']$, eq.(2.7), which subsumes all the
pre-exponential factors a la ref [4], is
\begin{eqnarray}\label{2.13}
F_{FV}[q,q']&=& \exp \{- S_{FV}[q, q']/\hbar\} \nonumber \\
S_{FV}[q, q']         &=& \frac{1}{2}\sum_\alpha \frac{c_\alpha^2(q-q')^2}{\hbar m_\alpha\omega_\alpha^3}
\coth(\beta\hbar\omega_\alpha /2)(1-\cos\omega_\alpha t) \nonumber  \\
                      &-& \frac{i}{2}\sum_\alpha \frac{c_\alpha^2(q^2-q'^2)}{\hbar m_\alpha\omega_\alpha^3}
(t- \frac{\sin\omega_\alpha t}{\omega_\alpha})
\end{eqnarray}
From (2.13) we can identify the decoherence factor [5] as the first term $S^{(N)}_y$ of $S_{FV}[q, q']$ in (2.13),
which is seen to be $positive$.It can be recast as an integral over the "spectral density " defined by [4]
\begin{equation}\label{2.14}
J(\omega) \equiv \frac{\pi}{2} \sum_\alpha \frac{c_\alpha^2}{m_\alpha\omega_\alpha}\delta(\omega-\omega_\alpha)
\end{equation}
which allows it to be expressible as [5]
\begin{equation}\label{2.15}
S^{(N)}_y = \frac{1}{\pi}\int_0^\infty J(\omega)\frac{q_0^2}{\hbar\omega^2}\coth(\beta\hbar\omega /2)(1-\cos\omega t)
\end{equation}
where $q_0 = q-q'$ is the spatial separation of two different localized states [4]. The decoherence rate $\gamma_{decoh}$
is just the time derivative of $S^{(N)}_y/\hbar$, which simplifies in the regime of high temperature ($\beta = 1/k_B T$) for the
ohmic case $J(\omega)=\eta\omega$ as [5]
\begin{equation}\label{2.16}
\gamma_{decoh} = \frac{\eta q_0^2}{\pi\hbar^2\beta}\int_{-\infty}^{+\infty} d\omega\frac{ \omega \sin \omega t}
{\omega^2 + \omega_0^2}
\end{equation}
where we have inserted  an infrared cut-off frequency $\omega_0$
which, although not needed here, will be found useful for later
purposes (see below). We also note that the damping rate is
negligible compared with the decoherence rate [5]. The integral is
now evaluated as a contour integral to give
\begin{equation}\label{2.17}
\gamma_{decoh} = \frac{q_0^2\eta}{\beta\hbar^2}\exp\{-t\omega_0\}
\end{equation}
We now turn to a generalization of this 1D formula to 3D, so as to include the  effect of an external magnetic field.

\section{Effect of H-field on Decoherence }
\setcounter{equation}{0}
\renewcommand{\theequation}{3.\arabic{equation}}

The  generalization of the foregoing result to 3D is merely a matter of replacements like
$$ q q' \Rightarrow {\bf q}\cdot{\bf q'}; \quad  q x_{i\alpha} \Rightarrow {\bf q}\cdot{\bf x}_{i\alpha} $$
in the Hamiltonian (2.1) and the potential term (2.2). More interesting coupling structures arise with
the introduction of an external (magnetic) field. While pairwise interactions do not produce any new
structures, there is now the possibility of $three-way$ interactions which have more
interesting features. The simplest form of the latter for a spin-half magnet of moment $\mu \sigma$ is
\begin{equation}\label{3.1}
V_3 = \frac{q_0}{\hbar\omega_{i\alpha}} i\mu \sigma\times \textbf{H}\cdot c_\alpha\textbf{x}_{i\alpha}
\end{equation}
where the factor $i$ associated with $\mu \sigma$ meets the hermiticity requirement. And the factor in front
ensures dimensional homogeneity with the formalism of Section 2. Note the correspondence
$q \Leftrightarrow  \textbf{Q}$ where

\begin{equation}\label{3.2}
 \textbf{Q} \equiv \frac{q_0}{\hbar \omega_{i\alpha}} i\mu \sigma\times \textbf{H}
\end{equation}
The formalism now goes through just as in Section 2 (adapted to the 3D form) as indicated above, except for
the replacement of $q_0^2 = (q-q')^2$ in (2.15) by
\begin{equation}\label{3.3}
 \textbf{Q}^2 = \frac{q_0^2}{\hbar^2 \omega_{i\alpha}^2}\{i\}^2 (\mu)^2 2 H^2
\end{equation}
making use of the result $(\sigma\times H)^2$ = $2H^2$. Note that the presence of the factor$\{i\}^2$ in
(3.3) renders the quantity $negative$ as a whole, thus $reversing$ the sign of the decoherence factor!
This brings out the central result of this exercise: The 3-way coupling involving the magnetic field
gives a contribution to the decoherence with a $reversed$ sign. As to the magnitude of the effect
vis-a-vis the reference value (2.17), the counterpart of (2.16) reads as
\begin{eqnarray}\label{3.4}
\gamma_{decoh}^H &=& \frac{i^2 \eta q_0^2}{\pi\hbar^2\beta}\int_{-\infty}^{+\infty} d\omega
\frac{ 2\omega_H^2 \omega \sin \omega t}{(\omega^2 + \omega_0^2)^2} \nonumber \\
\omega_H^2       &=& (\mu )^2 H^2/ \hbar^2
\end{eqnarray}
where the effect of the infrared cut-off frequency appears (even more strongly) in the last factor.
Evaluation of the integral as a contour integration over the double pole finally gives
\begin{equation}\label{3.5}
\gamma_{decoh}^H = \frac{q_0^2\eta }{\beta\hbar^2}\frac{i^2 \omega_H^2 t}{\omega_0}\exp\{-t\omega_0\}
\end{equation}
The value of $t$ in front of the exponential on the rhs is $ t \sim \omega_0^{-1}$, so that the
ratio of the two decoherence factors (3.5) to (2.17) finally works out as
\begin{equation}\label{3.6}
\gamma_{decoh}^H /\gamma_{decoh} \approx i^2 \omega_H^2 / \omega_0^2
\end{equation}
The factor $i^2$ brings out the reverse sign, while the smallness of
$\omega_0$ wrt $\omega_H$ (note that $\omega_0$ is a small infrared
cut-off, while $\omega_H$ is a finite quantity) testifies to the
enhancement of the (coherence) effect.

\section{ Discussion}

This work was motivated by the need to study the role of a magnetic
field in the context of origin of life scenarios [8]. While
referring the interested reader for details of the precise mechanism
involved therein, suffice it to say that the principal object behind
the introduction of a magnetic field is its role in bringing about
$coherence$ in a system whether it is directed against the
dissipating effect of temperature or for generating an environment
conducive to a quantum scenario. Taking a simple algebraic
viewpoint, the respective signatures of coherence / decoherence must
be of $opposite$ signs in whatever definitions are employed for
their respective measures. Viewed in that light, the result (3.6) is
fully consistent with the purported introduction of a magnetic field
in the interaction of an open system with its environment, viz., to
reduce the effect of decoherence by a reversed sign of the
contribution of the magnetic field, taking advantage of a {\it
special 3-way coupling} involving the system, environment and the
(external) magnetic field. As to the relative magnitude of the
factor $\omega_H$ vis-a-vis  $\omega_0$ in Eq (3.6), the former,
being a finite quantity, easily exceeds the latter which is a small
infrared cut-off. Somewhat different approaches [9, 10] directed
towards avoiding decoherence  for open systems by adjusting external
controllable parameters (such as a magnetic field) seem to to
converge towards similar conclusions, although employing different
strategies.
\section{Acknowledgement}

I am grateful to Gargi Delmotte for extensive discussions; and to Dr J-J Delmotte for infrastructural support.

\section*{Appendix A. }

\setcounter{equation}{0}
\renewcommand{\theequation}{A.\arabic{equation}}

Here we outline some simplified expressions for  the quantities $\phi_\alpha$ and $\phi_\alpha'$ of Section 2, based
on an elementary derivation of the classical action for a harmonic oscillator in the presence of a $constant$ force $f$.
To that end we first put on record the more general result for  a time-dependent force $f(t)$, vide problem 3-11
of Feynman-Hibbs [6], mostly in their notation, from which the result for $f =const$ follows as a particular case.
[An added reason for this exercise is the appearance of an extra term in the classical action, over and above the
result listed in this classic book [7], in the hope of obtaining a verdict on its bona fides from interested experts].
The classical action for this problem is given by [7]
\begin{equation}\label{A.1}
S_{cl} = \int_0^T dt [\frac{1}{2} m \dot{x}^2 - \frac{1}{2} m \omega^2 x^2 + xf(t)]
\end{equation}
which can be transformed via the equation of motion as
\begin{equation}\label{A.2}
S_{cl} = \int_0^T dt [\frac{1}{2} xf(t)] + \frac{1}{2} m [ x_b \dot{x_b}- x_a \dot{x_a}]
\end{equation}
where $x_a$ and $x_b$ are the initial and final positions and their dots are the corresponding velocities.
Now the solution of the equation of motion may be written as
\begin{equation}\label{A.3}
x = c \exp{i\omega t}+c^* \exp{-i\omega t}+ \frac{1}{m} \frac{1}{D^2+\omega^2}f(t)
\end{equation}
where the last term -- the particular integral $PI$--is obtained through the following steps
\begin{eqnarray}\label{A.4}
 PI &=&  \frac{1}{D^2+ \omega^2}f(t)/m = \frac{1}{2im\omega}[ \frac{1}{D-i\omega}- \frac{1}{D+i\omega}]f(t) \nonumber \\
    &=&\frac{1}{2im\omega}e^{i\omega t}\int_0^t e^{-i\omega t'} f(t')dt' + c.c. \nonumber \\
    &=&\frac{1}{2im\omega}e^{i\omega t}F_\omega (t) -\frac{1}{2im\omega}e^{-i\omega t}F_\omega^* (t)
\end{eqnarray}
thus defining the integral function $F_\omega(t)$ and its complex conjugate $F_\omega^*(t)$.
The constants $c$ and $c^*$ are determined from $x_a = c + c^*$ and
$$ x_b = ce^{i\omega T}+ c^*e^{-i\omega T}+ \frac{1}{2im\omega}[e^{i\omega T} F_\omega(T)- e^{-i\omega T}F_\omega^*(T)]$$
Straightforward substitution in (A.2) and simplification gives
\begin{eqnarray}\label{A.5}
S_{cl}&=& \frac{m\omega}{2\sin\omega T}[(x_a^2+x_b^2)\cos\omega T - 2 x_a x_b + \frac{2x_b}{m\omega}
\int_0^T f(t) \sin\omega t dt \nonumber \\
      &+& \frac{2x_a}{m\omega}\int_0^T f(t) \sin\omega (T-t)dt-\frac{2}{m^2\omega^2}
      \\
      & &\times\int_0^T dt \int_0^t ds f(t)f(s)\sin\omega(T-t)\sin\omega s] + S(EXTRA)
\end{eqnarray}
which agrees with Feynman-Hibbs [7]as well as Weiss [5], $except$  for the term  $ S(EXTRA) $ given by
\begin{equation}\label{A.6}
S(EXTRA) = \frac{1}{2m\omega}\int_0^T dt  \int_0^t ds f(t) f(s) \sin\omega(t-s)
\end{equation}
Unfortunately we are unable to write off the last term, eq (A.6), from the complete solution given in [7].
We  also note that it does not figure in Weiss [4]either, although we could not detect any error
in our (repeated) calculations. Its source can be traced to the integral term in (A.2), wherein the complementary
part of the $x$-solution, eq (A.3), contributes an amount
$$ \frac{1}{2}[ c F_\omega^*(T) + c^* F_\omega (T)] $$
which remains after identifying all the terms of prob 3-10 in  Feynman-Hibbs [7].

\end{document}